# Electric control of antiferromagnets

I. Fina[1,2], X. Marti[2,3]

[1] Institut de Ciència de Materials de Barcelona (ICMAB-CSIC), Campus UAB, Bellaterra 08193, Barcelona, Spain

[2] IGSresearch, C/ La Coma, Nave 8, 43140 La Pobla de Mafumet (Tarragona), Spain

[3] Institute of Physics ASCR, v.v.i., Cukrovarnicka 10, 162 53 Praha 6, Czech Republic

Contact: ignasifinamartinez@gmail.com, xavi.marti@igsresearch.com

I. F. and X. M. contributed equally.

## Abstract

In the past five years, most of the paradigmatic concepts employed in spintronics have been replicated substituting ferromagnets by antiferromagnets in critical parts of the devices. The numerous research efforts directed to manipulate and probe the magnetic moments in antiferromagnets have been gradually established a new and independent field known as antiferromagnetic spintronics. In this paper, we focus on the electrical control and detection of antiferromagnetic moments at a constant temperature. We address separately the experimental results concerning insulating and metallic thin films as they correspond to voltage and electrical current controlled devices, respectively. First, we present results on the voltage control of antiferromagnetic order in insulating thin films. The experiments show that voltage pulses can switch the chirality of a modulated antiferromagnetic structure. Second, we describe the recent advances in metallic antiferromagnetic systems. We present results obtained with the first USB-operated portable device able to perform the non-volatile electrical current-induced switching of an antiferromagnet combined with magnetoresistive readout at room temperature. We discuss on potential applications that can be realized using antiferromagnetic memory cells.

## Introduction

Antiferromagnetic materials concentrate numerous features that encourage the research on its integration in devices for both information processing and storage: i) antiferromagnetically-coupled moments are not altered by moderate magnetic fields, ii) the absence of dipolar interactions allow the stability of multiple magnetic configurations, iii) the nearly null stray fields reduces the cross-talk among neighboring elements within the same device, iv) they are more abundant in nature than ferromagnets and generally display higher magnetic ordering temperatures. While this list of features is noteworthy, antiferromagnetic materials have remained away from intensive research for many years due to the difficulties in manipulating and probing their magnetic order. Indeed, Louis Neel literally defined antiferromagnetic materials as "interesting but useless" during his Nobel lecture in 1970. Despite not experiencing a furious beginning in the applications front, research on the integration of antiferromagnetic materials in ferromagnetic-inspired devices has been important [1-4]. Works based on spintronic functionalities characterization on antiferromagnets have grown exponentially during the last decade as reviewed in other works [5-7]. Nowadays it is possible the control and detection of antiferromagnetic moments using electric pulses, which will be the topic covered in the present work.

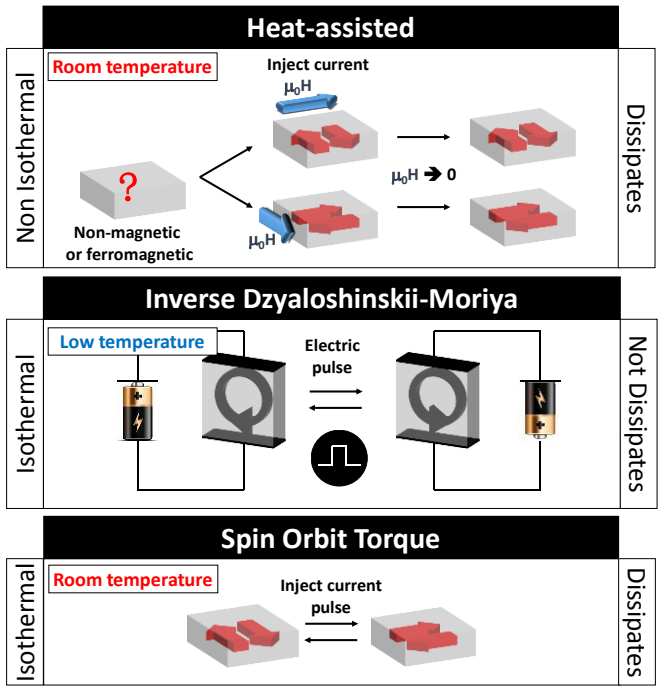

*Figure 1. Summary of the three mechanisms that allow electric control of magnetic order in antiferromagnetic materials. On the left, it is indicated if the process is isothermal. On the right, it is indicated if the process dissipates energy by Joule heating effect.*

We show the footsteps of both the control and the monitoring of antiferromagnetic ordering covering from the first experiments based on the combination of magnetic field and temperature ramps until the recent experiments in which the magnetic state is only manipulated by electrical means. We cover two sets of experiments where the control the antiferromagnetic order is achieved by means only of electric pulses of voltage or current. We also show results obtained by a prototype device of a memory-cell which is capable of editing and retrieving information from an antiferromagnet using electrical current pulses at room temperature. Finally, we discuss about some out of data storage potential applications of antiferromagnets.

## Preliminary experiments

The very first experiment showing the manipulation and readout of antiferromagnetic ordering was contemporaneous with the Nobel Prize awarded to Louis Neel. In 1971, W.B. Muir and J.B. Strom-Olsen observed that the electrical resistance of a single crystal of Chromium depends on the way how the sample was magnetically *prepared*: if the sample was heated above the Neel temperature ($T_N$~37ºC) and field-cooled, then the electrical resistance would change by ≈2% depending on the direction of the magnetic field while cooling down [8]. It is worth noting that the Neel noted in his Nobel lecture that ferromagnets and antiferromagnets would be equivalent for effects that are an even function of the magnetic moment and, remarkably, the ultimate origin of the electrical resistance difference observed in Chromium, namely the anisotropic magnetoresistance, which was at the same time being studied in ferromagnets by T. McGuire and R. Potter at IBM research laboratories[9] and was found to be an even function of the magnetic moment. On these grounds, the first integrated circuits using anisotropic magnetoresistance for the magnetic data readout would not see any significant difference between ferromagnets and antiferromagnets but ferromagnets were much easier to manipulate, making them the only option for applications.

More recently, the pioneering field-cooling experiments by W.B. Muir and J.B. Strom-Olsen have been revisited under a variety of sample geometries. The exchange-spring formed at the interface between a ferromagnet and antiferromagnet was employed to manipulate the latter in a tunnel junction [10, 11]. An ohmic-like vertical device was reported also inspired on the exchange-spring control of an antiferromagnet [12]. D. Petti and co-workers produced the first thermally-assisted tunneling device with antiferromagnetic electrodes without the need of any auxiliary ferromagnetic layer [13]. Marti and Moriyama returned to the seminal ohmic readout strategy based on anisotropic magnetoresistance in single-layer devices made of FeRh alloys (figure 1), now switching reversibly at room temperature [14, 15]. In this direction, D. Kriegner shed light on one subtle property of antiferromagnets which is not matched by antiferromagnets: the absence of dipolar interaction within the bulk of the memory element allows to lock multiple intermediate configurations between two opposite saturation states in MnTe [16]. All these papers, however, rely on the combination of heat and large magnetic fields or subsidiary ferromagnets for editing the magnetic configuration of the antiferromagnet thus setting a considerable distance to practical applications.

One branch of the search for strategies to circumvent the need of thermally-assisted methods emerged from the realm of multiferroic magnetoelectric materials. It is in this research area where the first experimental demonstration of electrically controlled antiferromagnetic moments in epitaxial thin films took place. Thanks to the tight interplay between electrical polarization, crystal structure distortion and magnetic ground state, in presence of strong spin-orbit coupling, one can directly control the orientation of magnetic moments using electric voltage pulses as we shall cover in section III (figure 1). Also exploiting the spin-orbit coupling, but in this case requiring electrical currents a new wave of phenomena emerged two decades ago to control ferromagnetic moments. The list of effects comprises Spin Transfer Torque, Spin Hall Effect, and, more recently, Spin-Orbit Torque [17]. The latter was employed to demonstrate the all-electrical isothermal manipulation of antiferromagnetic moments at room temperature using current pulses (figure 1). Section IV covers this experimental realization and, in particular, the development of a tabletop USB-controlled electronic board that can switch antiferromagnetic moments at room temperature and, from the theoretical perspective, at much greater speeds than it is nowadays possible in ferromagnets.

# Voltage switching of an antiferromagnet

## Background

Magnetic ferroelectric attracted a lot of interest due to its large magnetoelectric coupling. The large coupling in these materials results from the fact that ferroelectric order stabilizes owing to the presence of particular magnetic order. The chimeric example of this class of materials is TbMnO3 (TMO). In 2003, Tokura's group observed the presence of ferroelectric order in TMO, plus the largest up to that date observed magnetoelectric effect [18]. In fact, they observed the full control of ferroelectric order by magnetic field. Other seminal works characterizing parent compounds set down a family of materials were similar phenomenon was observed. Afterwards, neutron diffraction experiments [19] allowed to resolve that the antiferromagnetic nature of TMO below $T_N$=40K. At the so-called lock-in temperature ($T_{lock-in}$=27K) a cycloidal phase set in. Cycloidal magnets will be those focus here, although other classes of magnetic ferroelectrics exist (see their classification [20]). In cycloidal magnets the adjacent spins are rotated by a certain angle resulting from magnetic frustration. Cycloidal magnets are an intermediate phase between ferromagnetic order (where the angle between adjacent spins is 0°) and the collinear antiferromagnetic one (where the angle between adjacent spins is 180°).

Owing to that the magnetic structure in this class of materials was revealed, theory started to be developed. The most extended description that explains the observed large coupling is the Inverse Dzyaloshinskii-Moriya Effect (DME) [21-23]. DME is present only if important spin-orbit coupling exists. In the archetypical cycloidal REMnO3, the cycloidal spin arrangement results in a coherent displacement of the oxygen atom ($?_{ij}$) with respect the adjacent manganese one following the expression $?_{ij} = A_{SO} \cdot S_i \times S_j$, where subscripts indicate i and j adjacent manganese atoms with **S** spin orientation and $A_{SO}$ is a coefficient that must depend on Spin-Orbit coupling. DME in insulators results in the appearance of polarization, a surface charge. The sign of the surface charge is defined by the mentioned displacement, and therefore by the magnetic order and the chirality of the spins ($S_i \times S_j$)[1]. The cycloidal plane (the one containing the projection of the spins that rotate across the lattice) can be manipulated by the application of large magnetic fields, as it was well-known for previously studied cycloidal (helicoidal) magnets. In particular, the cycloidal plane rotates by 90° by application of magnetic field (flop). By the same token, the polarization also rotates by 90º, resulting in the observed large magnetoelectric coupling.

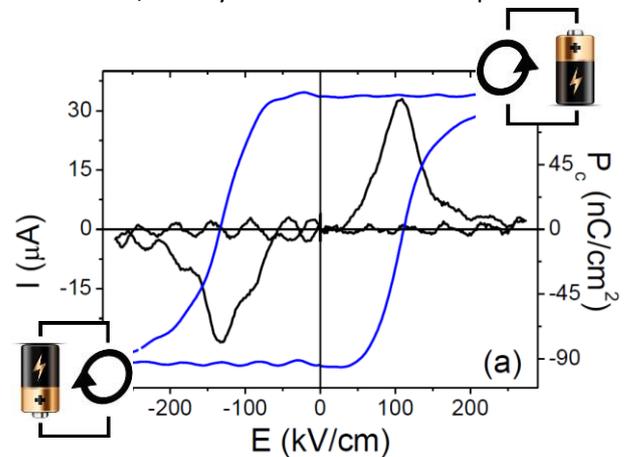

Figure 2. Ferroelectric hysteresis loop of an orthorhombic YMnO3 film. On the left axes, the measured current, where one can infer the two ferroelectric current switching peaks and, on the right axes, the ferroelectric polarization. The two different polar states correspond to the two different signs of the chirality of the antiferromagnetic order. Adapted from [25].

---

[1] Spin current theory described in ref.: [H. Katsura, N. Nagaosa, and A. V. Balatsky, "Spin current and magnetoelectric effect in noncollinear magnets," Physical review letters, vol. 95, p. 057205, 2005] explains, as well, the magnetoelectric phenomena in cycloidal magnets.

## Experimental results

Indeed, the surface charge can be switchable by the application of an external electric field, which in fact implies that the material is ferroelectric. The possibility to switch the ferroelectric polarization in cycloidal magnets was already observed in ref. [18]; however, the ferroelectric switch was not done isothermally, and required to cross the $T_{lock-in}$ while cooling the materials under the application external electric field. Afterwards, in DyMnO$_3$ [24] crystals and in YMnO$_3$ [25] (figure 2) films (with less than 5V) was observed the isothermal switch of ferroelectric polarization.

The isothermal electric switch of polarization in cycloidal magnets implies the change of sign of the chirality magnetic order of the magnetic ferroelectric; thus, the electric manipulation of the antiferromagnetic order (as sketched in figure 2). Comparing with the mentioned electrical current assisted techniques described in previous sections, the replacement of the electric field by the electric current implies the possibility to manipulate magnetism without the dissipation of heat concomitant to the presence of current. The works that have reported on the isothermal switch of polarization and chirality by electric field are nowadays abundant, which opens the door to the study of the electric field manipulation on a large number of antiferromagnetic materials. The fact that cycloidal magnetic order remember chirality sign after cycloidal flop or by increasing the temperature much above $T_N$ ensures that cycloidal magnets fulfill the robustness characteristic that makes antiferromagnets attractive for applications [26-28].

However, as mentioned, cycloidal magnetic order only exists at very low temperature. On this regard, BiFeO$_3$ can show up as an alternative. Although, it is not strictly speaking ferroelectric magnet, because ferroelectricity does not result from magnetic order, it is antiferromagnetic and ferroelectric. It shows cycloidal magnetic order [29-31] and is has been reported that there is coupling between the cycloidal order and the electric field [32-34]. This coupling has been used for probing the manipulation of magnetic order in multiferroic heterostructures based on BFO [35]. However, there is a lack of information regarding the potentiality of BiFeO$_3$ for applications without the requirement of any subsidiary ferromagnet, except the characterization of the antiferromagnetic/ferroelectric BiFeO$_3$ domain walls (ferromagnetic in nature) by the measurement of magnetoresistance and anisotropic magnetoresistance [36, 37].

# Current switching of an antiferromagnet

## Theoretical background

Electrical currents that flow across a solid interact with the stationary atoms in a number of ways. In particular, the relativistic spin-orbit coupling interaction produces a magnetic field acting on each atomic site. The theoretical proposal was published in 2014 by J. Zelezny [38] and we summarize some key aspects in the following. While the magnitude and direction of such magnetic field has to be calculated for each specific atomic topology (atoms, angles and distances), there are symmetry rules that apply. The most critical one is that to have a non-zero effect on a particular lattice site, there must be broken inversion symmetry. There are additional rules that are of interest in the present case. For instance, if the electrical current direction is reversed the sign of the generated magnetic field changes. Also, if the atomic environment of the central atom is mirrored the magnetic field changes sign. On these grounds, any crystal structure in which the chemical environment around the magnetic moments is alternated, will experiment an alternate magnetic torque on each magnetic moment [see sketch in figure 3(a,b) for Mn$_2$Au]. Some privileged crystal structures have a chemical modulation that matches the magnetic modulation. For instance, since in tetragonal CuMnAs the first half of the unit cell and the second half are mirrored respect to the Mn sites (which carry the magnetic moments), the net effect of an electrical current flowing along the basal plane is the creation of an alternated magnetic field pattern along the entire thin film volume. Due to this spatially modulated torque, the staggered Mn magnetic moments can synchronously rotate in the plane and eventually get stabilized along particular directions according to the magnetic anisotropies present in the crystal. This effect is not exclusive of the CuMnAs lattice and is expected to occur also in other antiferromagnetic materials with high $T_N$ such as $Mn_2Au$ [38].

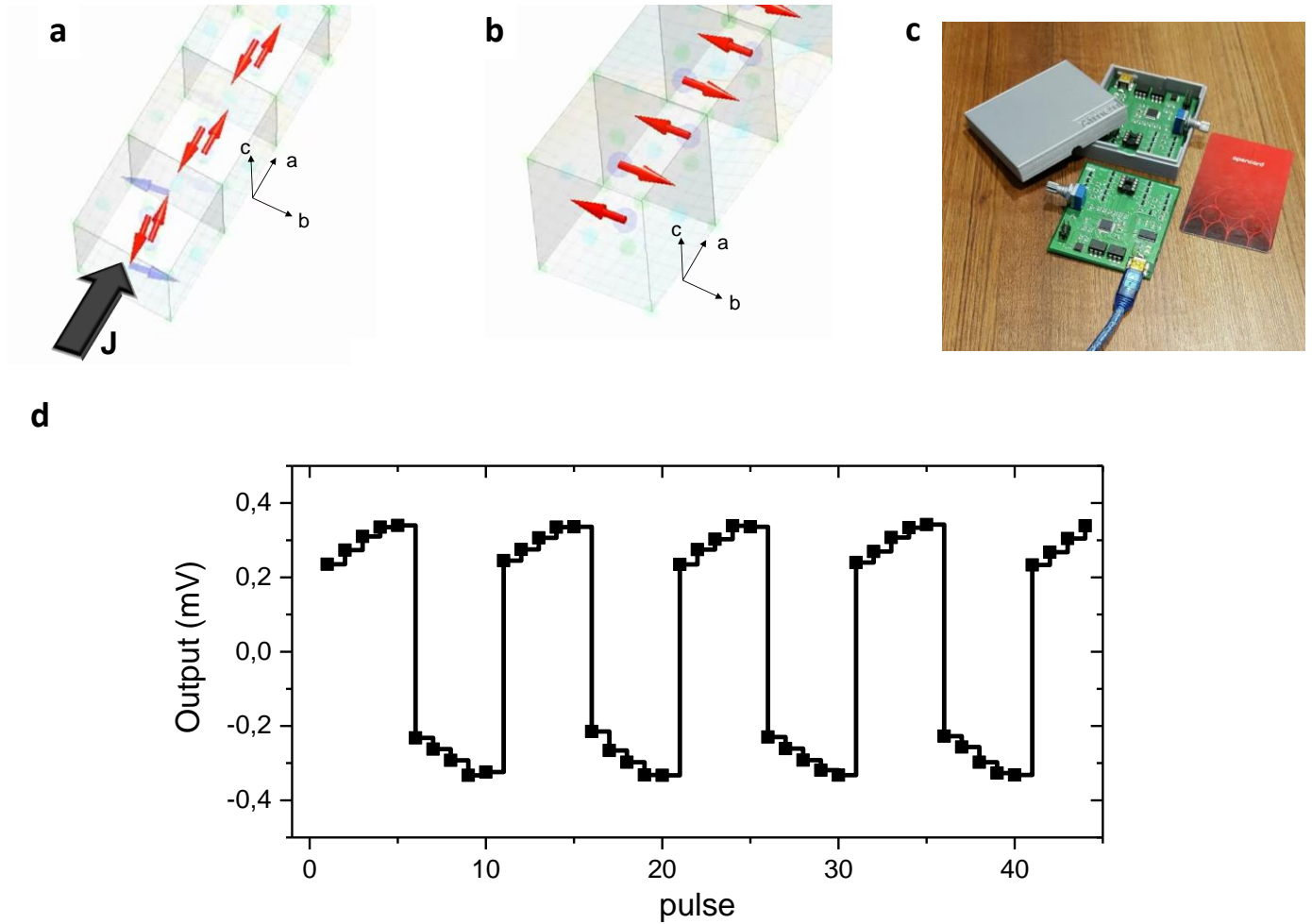

Figure 3. (a,b) Theoretical model for the electrical switching of an antiferromagnet. The electrical current pulse J (green arrow) flows along either the [100] or the [010] direction of CuMnAs. (a) The Mn magnetic moments rotate synchronously in opposite directions and, after the pulse, (b) the moments have rotated by 90º so they point perpendicular to the current. In reality, the electrical switching manifests via the motion of the domain walls which are favorable to the torques delivered by the current pulse. During the readout, implemented by an anisotropic magnetoresistance measurement, the electric current pulse scattering probes the population of each type of antiferromagnetic domain in the sample. (c) USB-device used to switch magnetic order in a CuMnAs film as shown in figure (d). (d) USB-device output for 5 consecutive pulses of the same polarity repeated several time for alternating polarities. Equivalent to the results show in [39].

Experimental results

The antiferromagnetic spin-orbit torque switching has recently been experimentally demonstrated in c-oriented epitaxial thin films of tetragonal CuMnAs [39]. When electrical pulses with current densities of the order of $10^{-7} Acm^{-2}$ are driven along the [100] or [010] directions the Mn magnetic moments rotate so that they point perpendicular to the current in consistency with the theoretical calculations. Electron microscopy studies have revealed that this idealized picture is taking place preferentially at the adjacent areas of those antiferromagnetic domain walls which are favorable to the spin-orbit torque delivered by the writing current pulses.

In order to implement the electrical switching in a compact device so that writing and reading can be executed sequentially a CuMnAs cross-bar structure needs to be patterned so that two orthogonal branches are used to switch the antiferromagnetic moments by 90 degrees roles in an alternate sequence. The reading is performed by measuring the electrical resistance of a CuMnAs bar. Thanks to the anisotropic magnetoresistance, the relative orientation of the staggered magnetic moments respect to the electrical current direction renders distinct stable and switchable memory states.

Of relevance here, the CuMnAs devices have the ability to incrementally transit from two opposite saturating resistance states if shorter pulses are applied. Each individual pulse assists the gradual expansion of the antiferromagnetic domain walls favorable to the current pulse direction. This ability offers the possibility to employ a CuMnAs device as a passive incremental counter as we shall discuss in section V.

### USB-device switching antiferromagnetic memory cell

The operation of the antiferromagnetic memory cell requires the application of current pulses along orthogonal directions for switching the magnetic state. Additionally, information can be read at any moment via anisotropic magnetoresistance by injecting a small amount of current along a given direction and measuring its electrical resistance. Since the entire experiment occurs at room temperature, the actual list of laboratory equipment is fairly reduced compared to the pioneering field-cooling experiments: a current source, a multiplexer to drive the current along selected branches of the cross-device, and a voltmeter. All such components exist as surface mounted devices and can be easily integrated into a single credit-card sized compact printed circuit board [shown in figure 3(c)]. We realized this board using a 16-bit analog-digital converter and a standard microcontroller. The communication of the board with the external devices (computer, smartphone, etc.) is achieved via a USB socket which, in turn, delivers the necessary power to switch the antiferromagnetic domain configuration. The multiplexing task is performed by a set of transistors that are controlled by the microcontroller which can open a particular current path with a rise time of 5 microseconds.

## Short-term applications

### Incremental non-volatile counter with a single-bit

One of the most employed data processing techniques in modern electronics is the incremental counter. Having the ability to monotonically count is crucial for complex tasks such as managing the road traffic or for many simple tasks as counting the number of people that has entered an airplane. The device that has to implement this function needs a non-volatile memory element, a method to increment a known amount sequentially, and a method for resetting the counter to zero. All these tasks are naturally implemented in the CuMnAs bars that we have presented in the previous section [as shown by figure 3(d)].

The electrical current pulses required for the spin-orbit torque switching of the CuMnAs bar can be obtained by a number of ways. Given the longitudinal ohmic resistance of a bar of the aforementioned CuMnAs device, 5V would suffice to achieve the electrical current densities necessary to realize the electrical switching. While 5V can be obtained by a standard battery or current source, a commercial piezoelectric cable could deliver this voltage pulse upon the pressure of a vehicle travelling on top of a coaxial piezoelectric cable.

By placing a voltage divider connecting both ends of the piezoelectric cable, one can tune the voltage pulse stemming from the piezoelectric cable and produce smaller amplitude pulses so that the antiferromagnetic device operates as an incremental counter of events triggered by mechanical pressure. When the electrical resistance of the bar would approach the opposite saturation limit a perpendicular pulse would reset the counter by resetting the antiferromagnetic moments in the initial configuration. On these grounds, the complete set of a renewable electrical generation and a non-volatile counting integrated in a single bar with no additional electronic components produces a completely passive device. Of course, some active elements need to be employed to extract the information from the antiferromagnetic memory device. However, if the counting method is implemented by a passive and non-volatile CuMnAs microdevice, the frequency at which this information is retrieved by the active elements can be much lower and thus the energy necessary to operate this counter is reduced. The synergy of lower energy consumption and a passive functionality is a critical for the successful deployment of massive networks of small devices in charge of small tasks but covering a very large geographical area. It is worth noting that many of the memory chips employed in basic sensor boards employed nowadays in the frame of smartcities demand very small amounts of memory and a very limited set of functions and counting is a central one.

It is perfectly possible to implement both proposed applications based on antiferromagnets by a number of alternative materials, components and strategies. Here, we aimed demonstrating two embodiments that can be realized with the already existing sets of samples. These short-term applications can play a critical role in delivering the necessary independent testing of the basic functionalities (for instance, information caching or counting) while long-term projects such the integration of antiferromagnets in large-capacity memory devices is researched.

### Caching and self-encryption of magnetic patterns

From technological point of view magnetic media is one of the prime supports for all sorts of long-term data storage [40]. There is an increasing interest for protecting the contained information from either sabotage or unauthorized access emerged and would not stop gaining attention in the present context of information technologies[41]. In fact in a recent report by the SIA and SRC [42], both associations semiconductor industry, it is stated the interest on that the material containing the information is the same that is able to protect the data, before the use of any encrypting algorithm. On this line, it is very appealing to turn magnetic objects invisible despite it has been shown in devices restricted to cryogenic temperatures.

Antiferromagnetic materials can accomplish with these requirements due to they are magnetically invisible to the nowadays commercial reading devices. The same thermal excursion across the ferromagnetic to antiferromagnetic phase transition that has been employed for imprinting a specific domain configuration in FeRh thin layers, can also be used to store ferromagnetic information, which is spontaneously clocked at room temperature, where the material is antiferromagnetic. Due to that the ferromagnetic to antiferromagnetic phase transition in FeRh is link to an important change in the FeRh unit cell volume, in recent reports [43], it has been demonstrated that the writing and retrieve of magnetic information can not only be done by heat, also with the use of applied voltage to an stacked piezoelectric material (figure 4).

### Conclusion

Antiferromagnetic spintronics has traveled along the footsteps of ferromagnets in the past decade. On the strong points side, antiferromagnetic spintronics provides robustness against external magnetic field perturbations (up to 12 T could not perturbate the antiferromagnetic memory bits), the absence of stray fields that may interact with neighboring bits and a theoretically 3 order of magnitude higher switching speeds than in the case of ferromagnetic analogous blocks. On the weak aspects, antiferromagnetic spintronics is a very young technological proposal that is so far limited to a rather small list of candidate materials which encompass certain complexities in the growth and sample preparation. This weakness is being mitigated by the increasing number of research groups that devote efforts into this new field as it can be seen in the now stable presence of a focused symposia on antiferromagnetic spintronics in international conferences. It is worth noting here that antiferromagnetic spintronics has already achieved a significant goal in technology transfer: the ability to electrically switch the memory states at room temperature and read the resulting value with simple Ohmic resistance measurements allowed the presentation of hand-size USB-based memory demonstration devices. Despite its very low current capacity (~1-10 bits), it is remarkable how after the seminal theoretical proposal published in 2010, the antiferromagnetic spintronics field has delivered a compact USB-powered device in just 6 years. However, the fail or success of a new technology proposal depends on how the existing technologies leave some niche spaces for it to develop while it is not competitive. In this paper, we have presented two strategies to stimulate antiferromagnetic spintronics in short-term despite the tight competition of ferromagnetic based technologies: first, the ability to count sequentially and, second, the ability to cache or self-encrypt a magnetic pattern or message.

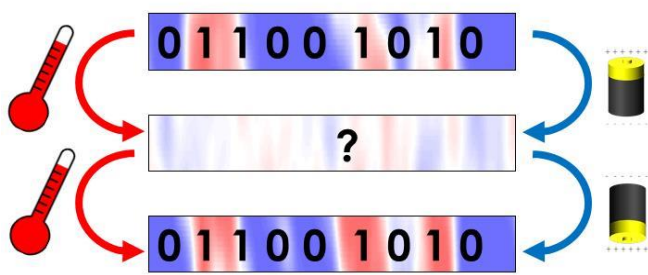

Figure 4. Sketch of the magnetic cloackable memory electrically and heat controlled.

# Acknowledgment

We acknowledge support from the EU ERC Advanced Grant No. 268066, from the Ministry of Education of the Czech Republic Grant No. LM2011026, from the Grant Agency of the Czech Republic Grant no. 14-37427, and from the Spanish Government Project no. MAT2015-73839-JIN. ICMAB-CSIC authors acknowledge financial support from the Spanish Ministry of Economy and Competitiveness, through the "Severo Ochoa" Programme for Centres of Excellence in R&D (SEV- 2015-0496). IF acknowledges Juan de la Cierva – Incorporación postdoctoral fellowship (IJCI-2014-19102) from the Spanish Ministry of Economy and Competitiveness of Spanish Government. "Institut de Català de Nanociència i Nanotecnologia" and "Col·legi Major Penyafort" are also acknowledged.